\documentclass[%
 preprint,
 amsmath,amssymb,
 aps,
pra,
]{revtex4-2}

\usepackage{graphicx}
\usepackage{dcolumn}
\usepackage{bm}
\usepackage{xcolor}
\usepackage{quantikz}

\usepackage[normalem]{ulem}
\usetikzlibrary{arrows.meta,positioning,calc}
\usepackage{hyperref}

\newcommand{\iu}{{i\mkern1mu}}

\begin{document}

\title{Efficient Tensor Decomposition of the Coulomb and Coulomb-like Potentials and Its Application to Probing Molecular Geometries Using a Quantum Computer}

\author{James Brown}
 \email{james.harry.brown@gmail.com}
\author{Masashi Takahashi\footnote{Present address: Brookhaven National Lab, Upton, NY, USA}}

\affiliation{qBraid Co.\\%
Chicago, IL\\
USA
}%

\author{Robert Wodraszka}
\affiliation{Queen's University\\
Kingston, ON \\
Canada}

\author{James Whitfield}
\email{James.D.Whitfield@dartmouth.edu}%
\affiliation{Dartmouth College \\
Hanover, NH\\
USA}

\date{\today}

\begin{abstract}
We present an efficient canonical polyadic decomposition of the Coulomb and Coulomb-like potentials in real-space grid representations suited to quantum computing applications. By decomposing the matrix elements of Coulomb-like potentials into low-rank Cartesian components, we achieve $\mathcal{O}(N^{1/3})$ T-gate scaling per Coulomb term, where $N$ is the number of grid points (basis functions) per electron, while maintaining high accuracy. This opens a path to significantly reducing the number of basis functions (i.e., qubits) required compared to previous methods by utilizing improved representations of the Coulomb potential that converge faster with grid spacing. The decomposition is also valid for certain pseudopotentials, which can decrease the cost significantly compared to previous work. This efficient decomposition enables the application of block-encoding techniques at larger grid spacings than with previous decompositions, and can therefore be used for algorithms that do not require high energy accuracy. We apply this decomposed Coulomb potential to hydrogen atom energy levels, the helium ground-state energy, and the Hartree-Fock potential energy curve of LiH, and show that modest basis sizes achieve semi-quantitative to quantitative accuracy. As an example of where these faster-converging Coulomb representations can be used on a quantum computer, we discuss their application to probing low-energy molecular geometries, and we estimate the resulting reduction in the cost of real-time evolution.
\end{abstract}

\maketitle

\section{\label{sec:intro}Introduction}

The equilibrium geometry of a molecule determines critical properties such as protein-ligand binding affinity and chemical reaction feasibility. However, finding the optimal geometry is computationally challenging, as the configuration space grows exponentially with the number of nuclei $N_\mathrm{nuc}$ in the molecule. The dimensionality of possible configurations is $3N_\mathrm{nuc}-6$, which when sampled n\"{a}ively with $L$ points per dimension yields $L^{3N_\mathrm{nuc}-6}$ possible configurations to evaluate. While quantum computers can in principle store exponential ($2^N$) information in $N$ qubits, loading this information requires exponential time, necessitating clever algorithms that can efficiently generate and probe desired regions of configuration space.

One popular technique to sample low-energy configurations classically is low-rank tensor factorizations\cite{White1992, Schollwock2011, Oseledets2011}. Density matrix renormalization group\cite{White1992, Schollwock2011} (DMRG) and its tensor-train generalizations\cite{Oseledets2011} compress the wavefunction representation through a tensor representation. These techniques also require the operators to be represented in tensor form. For example, tensor hypercontraction (THC)\cite{Hohenstein2012, Parrish2012} factorizes the electron repulsion integral (ERI) tensor into a product of two tensors. However, it has recently been shown that the effective rank of the decomposition must rise faster than linearly relative to the number of atomic orbitals.\cite{Oszmian_2026} This is because the parameter capacity of a strictly linear-rank CP decomposition is insufficient to cover the $\mathcal{O}(N^2)$ independent physical dimensions of electron repulsion integrals.\cite{Oszmian_2026} On quantum computers, CP decompositions reappear in the area of block encodings as similar structures can be exploited. Examples are the single, double, and THC factorizations of second-quantized Hamiltonians\cite{Lee2021} and, more recently, higher-order tensor factorizations for vibrational Hamiltonians\cite{kamakari2025higherordertensorfactorizations}.

Much of the work above is performed in second quantization, but it is expected that simulating electronic systems in a grid-like basis in first quantization will also be an efficient approach on quantum computers\cite{Babbush2023, Su2021}. In first quantization, each of the $\eta$ electrons is assigned to its own register of $\log_2 N$ qubits (where $N$ is the number of basis functions) encoding its position on the grid, and antisymmetry is enforced explicitly rather than through the exponentially large Fock space of second quantization. This representation is advantageous because the number of qubits scales as $\eta \log_2 N$ with the number of electrons $\eta$, and the kinetic energy operator acts locally on each register. However, the main bottleneck of utilizing this representation has been the application of the Coulomb potential in the block-encoding framework\cite{Su2021}. In this work, we address this bottleneck by presenting an efficient canonical polyadic (CP) decomposition of the Coulomb potential that achieves $\mathcal{O}(N^{1/3})$ scaling with substantially reduced circuit complexity. This improvement makes it feasible to probe molecular geometries for moderately sized molecules using quantum computers.

Our approach combines two key innovations: First, we directly decompose the Coulomb potential into Cartesian components using alternating least squares (ALS), (which can be thought of as a generalization of singular-value decomposition to more than two dimensions) achieving high accuracy with only rank-8 decomposition. Notably, this decomposition is performed only once and can be later utilized for any grid-spacing or molecular system. Second, we demonstrate how this efficient representation enables the exploration of molecular configuration space through quantum algorithms that can sample energetically favorable geometries.

Numerically, we compare several Coulomb representations on the hydrogen-atom spectrum with atomic units used throughout. More specifically, we show that low-lying energies of the Hydrogen atom can be retrained to $\sim 10^{-5}$~Hartree precision with only a rank-8 decomposition of a regularized Coulomb potential (Sec.~\ref{sec:hamiltonian}). We then compute the full configuration interaction (FCI) energy of helium on a $91^3$ grid with a CP-compressed wavefunction to chemical accuracy relative to the exact Born-Oppenheimer result. Next, we compute a Hartree-Fock potential energy curve for LiH and show that certain Coulomb-type potentials obtain a semi-quantitative accuracy much faster. Next, we show that the local part of a Li pseudopotential (Sec.~\ref{sec:cpdecomp}) can also be represented with a low-rank CP decomposition. We then obtain T-gate counts for the resulting block encoding with Qualtran (Sec.~\ref{sec:lcu}) to numerically verify the $\mathcal{O}\left(N^{1/3}\right)$ scaling. Finally, for a one-dimensional toy model, we discuss how a ground-state filter concentrates probability on low-energy nuclear geometries (Sec.~\ref{sec:geometry}).

Although our examples focus on ground-state filtering, the block encoding developed here can also be used for time-evolution. Qubitization-based Hamiltonian simulation and phase estimation both query the block encoding a number of times proportional to its normalization $\lambda$. A potential that converges at coarser grid spacing therefore lowers the cost per unit of simulated time. We give a back-of-the-envelope estimate of this benefit in Sec.~\ref{sec:timeevolution}.

\section{\label{sec:hamiltonian}The Hamiltonian in Discrete Variable Representation}

The Hamiltonian for our molecular system is defined using the discrete variable representation (DVR)\cite{Stare2003} as
\begin{equation}
    H = T+U+V
\end{equation}
where
\begin{equation}
\begin{array}{rl}
    T = & \sum_{i=1}^{\eta}QFT_i\left(\sum_{p\in G}\frac{n_p^2}{2}\ket{p}\bra{p}_i\right)QFT_i^{\dagger}, \\
    V = & -\sum_{i=1}^{\eta}\sum_{l=1}^{N_A}\zeta_l \sum_{p\in G}V_{p,l}\ket{p}\bra{p}_i, \\
    U = & +\frac{1}{2}\sum_{i\neq j=1}^{\eta}\sum_{pq\in G}V_{p,q}\ket{p}\bra{p}_i\ket{q}\bra{q}_j
\end{array}
\end{equation}
Here, $QFT$ represents the quantum Fourier transform, $p,q$ are multi-indices $p=(i_p,j_p,k_p)$ representing positions $x=i_p\Delta x, y=j_p\Delta x, z=k_p\Delta x$ with index range $G=\{i_p,j_p,k_p\in\{0,1,\dots,N^{1/3}-1\}\}$, so that there are $N^{1/3}$ grid points per dimension and $N$ in total, $\eta$ is the number of electrons, $N_A$ is the number of nuclei, with nucleus $l$ having charge $\zeta_l$, $V_{p,l}$ is the Coulomb kernel between grid point $p$ and nucleus $l$, and $\Delta x$ is the grid spacing. The quantum Fourier transform of the kinetic energy operator results in diagonal matrix elements in momentum space:
\begin{equation}
    T_{p, q} = \delta_{p, q}\frac{n_p^2}{2} = \delta_{p,q}\frac{n_{i_p}^2+n_{j_p}^2+n_{k_p}^2}{2},
\end{equation}
where $n_{i}=\frac{2\pi}{N^{1/3}\Delta x}\left(i-\frac{N^{1/3}}{2}\right)$ is the momentum along one axis for a box of side $L=N^{1/3}\Delta x$.

\subsection{Standard Coulomb Potential Representation}

In a standard DVR, the potential is diagonal with the potential evaluated at grid points. For the Coulomb potential, this takes the form
\begin{equation}\label{eq.standard}
    V_{p, q} = \frac{1}{\Delta x\sqrt{(i_p-i_q)^2+(j_p-j_q)^2+(k_p-k_q)^2}}
\end{equation}
This representation lends itself to circuits scaling as $\mathcal{O}(N^{1/3})$ where $N$ is the total number of basis functions per electron\cite{Su2021}. However, care must be taken at the point $i_p=i_q, j_p=j_q, k_p=k_q$ where the potential is undefined. A common approach is to add a regularization parameter $\epsilon$:
\begin{equation}\label{eq.regularized}
    V_{p, q} = \frac{1}{\Delta x\sqrt{(i_p-i_q)^2+(j_p-j_q)^2+(k_p-k_q)^2+\epsilon}}
\end{equation}
which eliminates the cusp singularity. Here $\epsilon$ is dimensionless and in the paper we use $\epsilon=10^{-4}$.

\subsection{Improved Coulomb Potential Representations}

Despite resolving the singularity, Eq.~(\ref{eq.regularized}) exhibits slow energy convergence for general grid-like bases\cite{Jones2016}. Several improved forms have been proposed. The representation of Ref.~\citenum{Lehtola2023} is designed such that the hydrogen atom ground state energy is exactly $-0.5$ Hartree. Another approach begins with a sinc DVR basis\cite{Colbert1992} and obtains a numerical representation by solving the Poisson equation using the DVR representation of the kinetic energy operator (the Green's function of the Laplacian)\cite{Jones2016}.

To compare these methods, we calculate the 1S, 2S, and 2P energies of the hydrogen atom with $\Delta x=0.6$ Bohr using Fourier Grid Hamiltonian basis functions:

\begin{table}[htbp]
    \centering
    \begin{tabular}{|c|c|c|c|}
    \hline
          & 1S  & 2P & 2S \\
         \hline
        Eq.~\ref{eq.regularized} & -0.4477 & -0.1250 & -0.1181 \\
        Ref.~\citenum{Lehtola2023} & -0.4845 & -0.1127 & -0.1329  \\
        Ref.~\citenum{Jones2016} & -0.4968 & -0.1250 & -0.1246\\
        Ref.~\citenum{Jones2016} CP rank 8 & -0.4968 & -0.1250 & -0.1246\\
        \hline
    \end{tabular}
    \caption{Energies (in Hartree) of hydrogen atom states for different Coulomb potential representations with grid spacing $\Delta x=0.6$ Bohr and 64 basis functions per dimension using a Fourier basis. The energy difference between the exact representation of Ref.~\citenum{Jones2016} and its rank-8 CP decomposition is on the order of $0.00001$ Hartree.}
    \label{tab:convergence}
\end{table}

As shown in Table~\ref{tab:convergence}, the energies are most uniformly accurate using Ref.~\citenum{Jones2016} (hereafter called the sinc potential). Therefore, we use this potential form for all subsequent results. The Ref.~\citenum{Lehtola2023} result deviates from $-0.5$ Hartree due to finite grid size; reducing the grid spacing to approximately $0.3$ Bohr yields the exact value. However, significantly more grid points are required to accurately represent the 2S/2P states due to their extended spatial tails.

\section{\label{sec:cpdecomp}Canonical Polyadic Decomposition}

\begin{figure}[htbp]
    \centering
    \includegraphics[width=\linewidth]{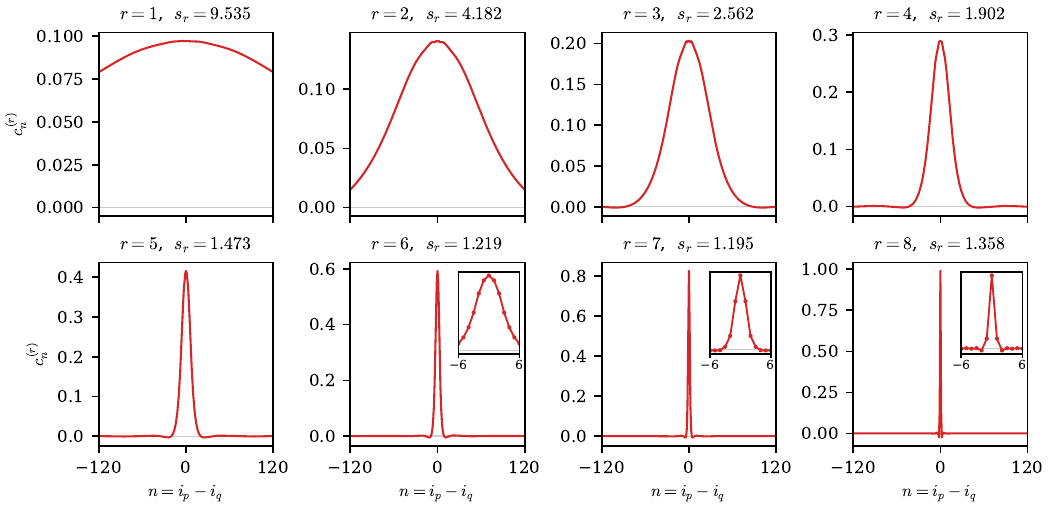}
    \caption{The eight factor vectors $c^{(r)}_n$ of the rank-8 decomposition in Eq.~(\ref{eq:cp_decomp}), over the full range of index differences $n=i_p-i_q\in[-120,120]$ used in the decomposition. The vertical axis is the dimensionless value of $c^{(r)}_n$, without the coefficient $s_r$, which is given above each panel. The $x$, $y$, and $z$ factors of each term are identical, so one is shown, and terms are ordered from broadest to narrowest. Insets show the grid points with $|n|\le 6$ for the three narrowest terms. Each term, multiplied by $s_r$ and tensored over $x$, $y$, $z$, contributes to the Coulomb kernel.}
    \label{fig:rank_8}
\end{figure}

Since the sinc potential coefficients lack exploitable special structure, we employ a different technique to achieve $\mathcal{O}(N^{1/3})$ scaling. We use a canonical polyadic (CP) decomposition\cite{Beylkin2002} to efficiently represent the Coulomb potential. The CP decomposition is a representation of a multidimensional array into a sum of rank-one tensors. This approach was first applied to quantum computing in Ref.~\citenum{Poirier2022}, which noted that the Coulomb potential can be represented as:
\begin{equation}
    \frac{\sqrt{\pi}}{2}\frac{1}{\sqrt{x^2+y^2+z^2}} = \int_{0}^{\infty}\mathrm{d}t \exp[-(x^2+y^2+z^2)t^2]=\int_{0}^{\infty}\mathrm{d}t \exp[-x^2t^2]\exp[-y^2t^2]\exp[-z^2t^2]
\end{equation}
Numerical approximation of this integral (e.g., through quadrature) reveals that the Coulomb potential naturally forms a CP tensor. While Ref.~\citenum{Poirier2022} required rank $\approx 50$, Ref.~\citenum{Brown2019} achieved electronic energy calculations with rank 15, motivating our search for even lower-rank decompositions. The CP decomposition of the Coulomb potential is shown in Fig.~\ref{fig:rank_8}.

\subsection{Alternating Least Squares Decomposition}

\begin{figure}[htbp]
    \centering
    \begin{tikzpicture}[
        box/.style={draw, rounded corners, minimum width=3.4cm, minimum height=1.0cm, align=center, font=\small},
        fixed/.style={text=gray},
        arr/.style={-{Stealth[length=2.2mm]}, thick}]
      \node[box] (init) at (0,2.2) {Initialize rank $R$:\\$X, Y, Z \in \mathbb{R}^{(2N^{1/3}-1)\times R}$};
      \node[box, fill=blue!8] (ux) at (-4.2,0) {Solve for $X$\\{\color{gray}$Y, Z$ fixed}};
      \node[box, fill=blue!8] (uy) at (0,0) {Solve for $Y$\\{\color{gray}$X, Z$ fixed}};
      \node[box, fill=blue!8] (uz) at (4.2,0) {Solve for $Z$\\{\color{gray}$X, Y$ fixed}};
      \node[box] (conv) at (0,-2.2) {Check $X = Y = Z,\quad$ $\epsilon<10^{-12}$\\(spherical symmetry)};
      \node[box] (grow) at (0,-4.4) {$R \leftarrow R+1$\\(stop at $R=8$)};
      \draw[arr] (init.west) -| (ux.north);
      \draw[arr] (ux) -- (uy);
      \draw[arr] (uy) -- (uz);
      \draw[arr] (uz.south) |- (conv.east);
      \draw[arr] (conv.west) -| node[pos=0.75, left, font=\scriptsize, align=right]{no:\\next sweep} (ux.south);
      \draw[arr] (conv.south) -- node[right, font=\scriptsize]{yes} (grow.north);
      \draw[arr] (grow.east) -- ++(4.5,0) |- (init.east);
    \end{tikzpicture}
    \caption{Schematic of the ALS sweep used to decompose the Coulomb kernel. Within a sweep, each Cartesian factor matrix is obtained in turn by a linear least-squares solve with the other two held fixed (blue boxes). Sweeps are repeated until the difference between the Cartesian components falls below $10^{-12}$. The rank is then increased by one and the procedure is repeated, restarting with a smaller $s_r$ for the new term if the $X=Y=Z$ symmetry is broken.}
    \label{fig:als_sweep}
\end{figure}
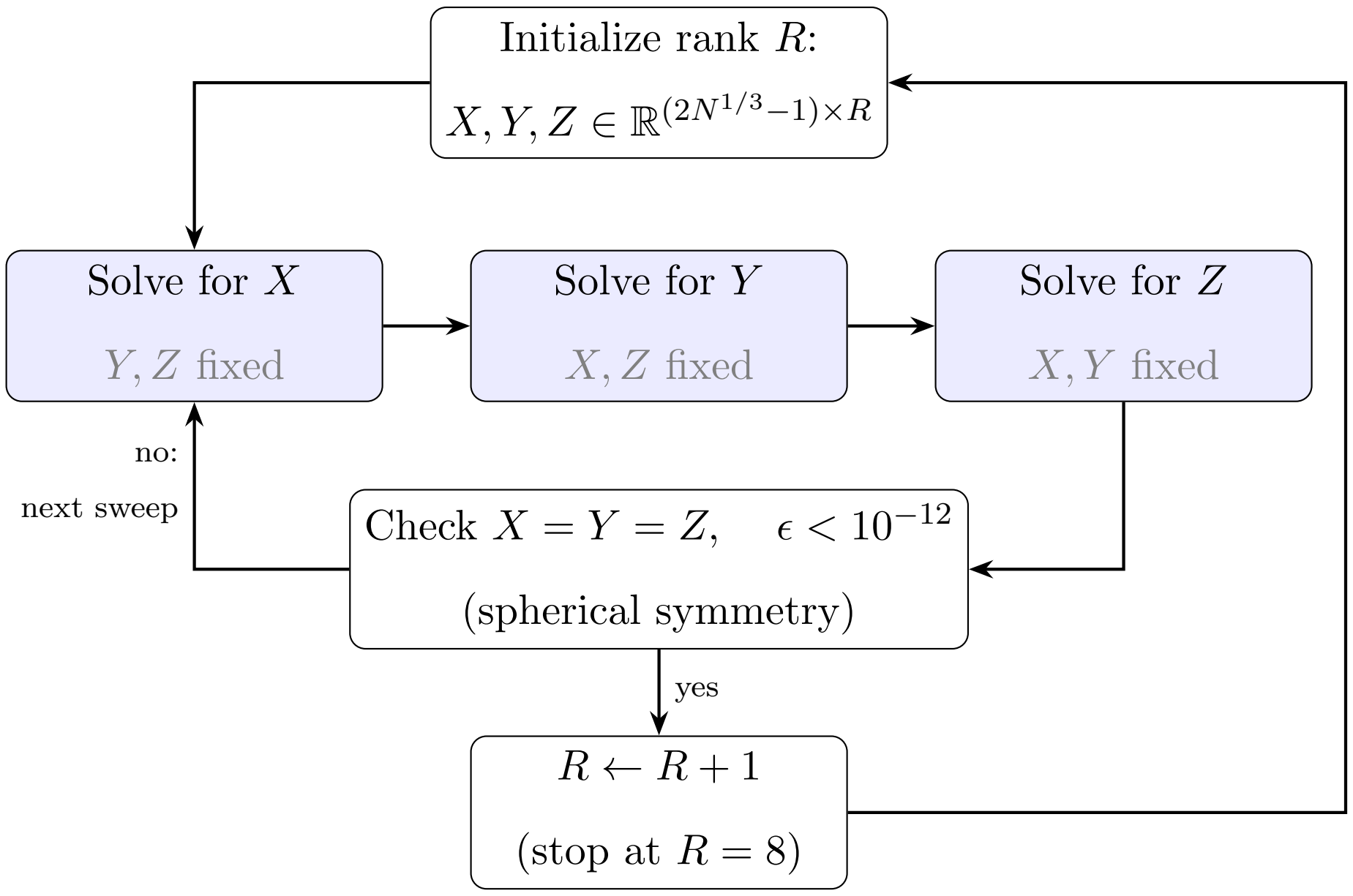

Unlike previous work, we directly apply alternating least squares (ALS) decomposition\cite{Beylkin2002} to the coefficients obtained from solving the Poisson equation in the sinc DVR. The decomposition is performed with respect to the $x, y, z$ coordinates. For a two-index array $M_{ij}$, the best rank-$R$ approximation $M\approx\sum_{r=1}^R s_r a^{(r)}_i b^{(r)}_j$ is given by the truncated SVD. This decomposition can also be obtained by solving the linear least-squares problem for $\{a^{(r)}\}$ with $\{b^{(r)}\}$ fixed, and vice versa, repeating until convergence. For tensors with three or more indices, there is no SVD algorithm that can be used to give a CP decomposition. However, the alternating least squares strategy still applies\cite{Kolda2009}. For a three-index tensor $V_{ijk}$ with factor matrices $X$, $Y$, $Z$ (columns $x^{(r)}$, $y^{(r)}$, $z^{(r)}$), the least squares problem (defined in Ref \citenum{Kolda2009}) is solved in order for $X$, $Y$ then $Z$. One cycle of the three updates is a \emph{sweep}, illustrated in Fig.~\ref{fig:als_sweep}.   While this is a computationally intensive non-linear optimization, we achieve excellent accuracy with only rank 8.

Our decomposition procedure:
\begin{enumerate}
    \item Begin with rank-3 decomposition and iterate until ALS converges
    \item Progressively increment the rank by 1
    \item Perform ALS decomposition to convergence threshold of $10^{-12}$
    \item Verify that each $x, y, z$ component (denoted $c$) remains equivalent for each term
    \item If symmetry breaks, restart ALS with a smaller coefficient for the added term
    \item Continue until rank 8 is achieved
\end{enumerate}

The resulting rank-8 decomposition yields energy differences less than $0.1$ millihartree for hydrogen atom 1S, 2P, and 2S states with grid spacing $\Delta x=0.6$ Bohr (see Table~\ref{tab:convergence}). A key feature of the converged decomposition is that, by construction of the ALS procedure, all three Cartesian factor vectors $c^{(r)}$ are identical for each rank-$r$ term. This reflects the spherical symmetry of the Coulomb potential and is verified at each step of the incremental rank increase. The electron-repulsion potential is represented as:
\begin{equation}\label{eq:cp_decomp}
    V_{p,q}=\frac{1}{\Delta x}\sum_{r=1}^8 s_r c_{i_p-i_q}^{(r)} c_{j_p-j_q}^{(r)}c_{k_p-k_q}^{(r)}
\end{equation}
where $s_r$ are positive real coefficients and there are eight $c^{(r)}$ vectors of length $2N^{1/3}-1$ that define the Coulomb potential.

For the one-electron potential with nuclei are at grid points, the same decomposition can be used:
\begin{equation}\label{eq:cp_one-body}
    V_{p}=\frac{Z}{\Delta x}\sum_{r=1}^8 s_r c_{i_p-R_i}^{(r)} c_{j_p-R_j}^{(r)}c_{k_p-R_k}^{(r)}
\end{equation}
where $(R_i,R_j,R_k)$ represent the location of the nuclei with charge $Z$. When nuclei are not positioned on grid points, the coefficients $c^{(r)}$ can be interpolated using the underlying sinc basis\cite{Jones2016}, eliminating the need for additional decompositions. 

Two things to note. First, the factors are only computed on index differences $|n|\le 120$ so this only covers grids with up to $N^{1/3}=121$ points per dimension. Second, the decomposition itself stays exact under a rescaling of $\Delta x$. Therefore, the decomposition need not be repeated for different grid spacings. In the future, a very large decomposition could be obtained and used repeatedly without more classical compute required.

A qualitative feature of the converged decomposition that is visible in Fig.~\ref{fig:rank_8} is that each of the terms resemble one another up to a rescaling of their width. This is consistent with exponential-sum (Gaussian-sum) approximations of $1/r$\cite{Beylkin2002, Braess2005, Beylkin2010}, in which the Gaussian exponents are distributed approximately uniformly on a logarithmic scale. In the future, these analytic expansions can possibly be used for two improvements. First, Gaussian factor vectors could be prepared\cite{poirier2021efficientevaluationexponentialgaussian} or applied with specialized circuits rather than generic alias sampling, which may lead to more T-gate-efficient block encodings. Second, a Gaussian-sum form provides a natural initial guess for ALS, and extending the decomposition to larger grids could be more efficiently obtained with the Gaussian starting guess. We leave both directions to future work.

\subsection{Relation to THC and to CP decompositions of the ERI tensor}

It is useful to compare the present decomposition with tensor hypercontraction (THC)\cite{Hohenstein2012, Parrish2012} which also uses the CP format and is now standard in quantum resource estimates\cite{Lee2021}. In THC, the ERI tensor in a standard quantum chemsitry orbital basis, is written as
\begin{equation}\label{eq:thc}
    (pq|rs) \approx \sum_{\mu,\nu=1}^{M} X_{p\mu}X_{q\mu}\,Z_{\mu\nu}\,X_{r\nu}X_{s\nu},
\end{equation}

where $X_{p\mu}$ are orbital values at a set of $M$ interpolation points (the rank of the decomposition) and $Z_{\mu\nu}$ is a central Coulomb-like matrix. In least-squares THC (LS-THC)\cite{Parrish2012}, $Z$ is obtained by a linear least-squares fit with $X$ fixed. Alternatively, all THC factors can be refined by alternating or nonlinear least squares\cite{Hummel2017, Lee2021}, as we do for the Coulomb potential here. An important difference is that THC factorizes the orbital indices with $M$ growing nearly linearly with system size. We instead factorize the Cartesian components of a system independent potential which has a rank that we argue grows logarithmically with system size. 

That being said, Oszmian and Lesiuk\cite{Oszmian_2026} recently showed that the effective CP rank of the ERI tensor cannot in general grow linearly with system size. They derived a lower bound $\propto N_{\mathrm{AO}}^2/\log_2^7 N_{\mathrm{AO}}$, where $N_{\mathrm{AO}}$ is the number of atomic orbitals. In an orbital basis, the ERI tensor depends on atomic positions and the atomic shapes directly. The $N_{\mathrm{AO}}$ scaling traces from the latter with ``the fundamental inability of a linear-rank global [CP] format to represent long-range interactions between orbital products in the regime where the multipole expansion is valid'' \cite{Oszmian_2026}. Using a grid, the size of the system enters the decomposition through the range of distances the representation must cover, i.e.\ the simulation box. In Ref \citenum{Beylkin2010}, it was shown that the number of terms grows only logarithmically with larger range for a fixed relative accuracy. Therefore, doubling the box size is expected to require only about one additional term. In quantum computer terms, the growth of the decomposition can be expected to expand by one for each additional qubit representing $x,y,z$.

\subsection{Many electron calculations}
\begin{figure}[htbp]
    \centering
    \includegraphics[width=0.9\linewidth]{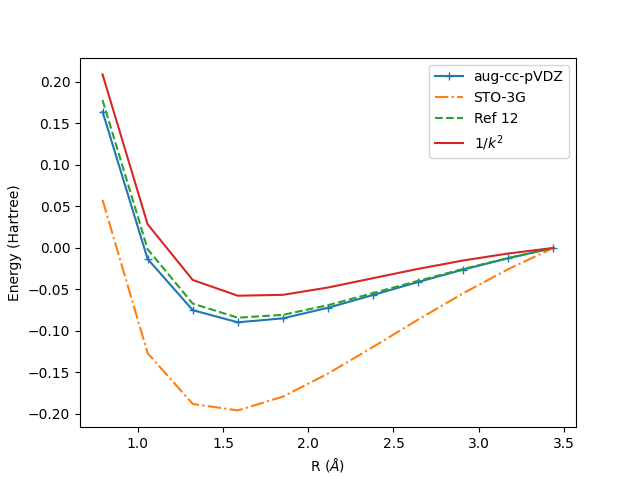}
    \caption{A Hartree-Fock calculation for LiH using a grid spacing of $0.5$ Bohr using the sinc DVR basis with both the Green's-function-derived Coulomb matrix elements from Ref.~\citenum{Jones2016} and the standard $1/k^2$ momentum space matrix elements. For comparison, calculations using standard Gaussian minimal STO-3G and double-zeta aug-cc-pVDZ basis are also presented. The standard $1/k^2$ potential shows a shallower minimum while the Ref.~\citenum{Jones2016} is qualitatively closest to the reasonably accuate aug-cc-pVDZ basis.}
    \label{fig:lihp5}
\end{figure}

To assess the accuracy of the potential derived above for multi-electron systems, we perform a Hartree-Fock calculation. The system we study is LiH with the matrix elements calculated as in Ref \cite{Jones2016}. The application of the Hartree-Fock operator has $\mathcal{O}(N \log(N))$ scaling for both the potential and kinetic energy operator terms due to $V_{pq}$ being triple-Toeplitz\cite{Jones2016}, meaning that each of the three Cartesian index differences $(i_p-i_q, j_p-j_q, k_p-k_q)$ enters independently. This structure permits fast matrix-vector products via three nested one-dimensional FFTs, enabling efficient classical pre-computation of the Fock matrix. As can be seen in Fig \ref{fig:lihp5} for the LiH potential energy curve with a grid spacing of $0.5$ Bohr, the sinc derived potential of Ref~\citenum{Jones2016} is qualitatively much closer to the reasonably accurate aug-cc-pVDZ basis than it is to the minimal STO-3G basis. The $1/k^2$ potential has a minimum that is shallower and is at a greater distance $R$. This means that the potential of Ref~\citenum{Jones2016} can provide reasonable representations of the potential energy surface with large grid spacing than smaller standard basis sets and other potentials. 

\subsubsection{Full electron calculation with CP decomposed form}
We can also examine the energy convergence of the potential with a full configuration interaction (CI) 2-electron calculation for the Helium atom using a CP decomposition of the wavefunction itself. The wavefunction is decomposed into tensors of each of the Cartesian coordinates but not between electrons, so that electron correlations within each Cartesian axis are captured while the inter-Cartesian coupling is handled through the CP rank.
\begin{equation}
    \psi(x_1,x_2,y_1,y_2,z_1,z_2) \approx \sum_{i=1}^{R_{\psi}}w_i f_i(x_1,x_2) f_i(y_1,y_2)f_i(z_1,z_2)
\end{equation}
where each $f_i$ is a matrix of size $N^{1/3}\times N^{1/3}$ and $R_{\psi}$ is the wavefunction rank, i.e., the number of terms in the expansion. The grid spacing is $0.1$ Bohr with $91$ grid points per direction, so each $f_i$ contains $91^2$ elements. Note that this is not a CP decomposition over the six individual coordinates. Instead, each factor $f_i$ depends on two coordinates (the same Cartesian component of both electrons) so the expansion is a CP decomposition of the order-3 tensor obtained by grouping $(x_1,x_2)$, $(y_1,y_2)$, and $(z_1,z_2)$ into single modes. Correlation between the two electrons along each axis is thus represented exactly within a factor, and only the coupling between different axes is limited by the rank $R_\psi$.

\begin{figure}[htbp]
    \centering
    \includegraphics[width=0.9\linewidth]{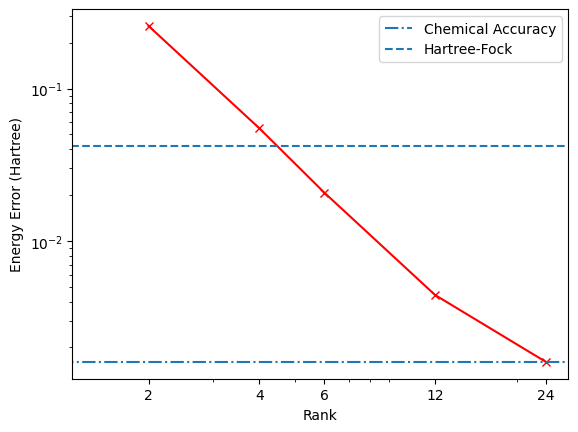}
    \caption{Using a CP representation (with respect to Cartesian coordinates and Hamiltonian) to calculate the ground state energy of Helium. With only a rank of 24, the ground state can be calculated to chemical accuracy. Hartree-Fock accuracy can be obtained with a rank of 6. Errors are measured relative to the exact nonrelativistic ground-state energy of helium ($-2.903724$~Hartree).}
    \label{fig:placeholder}
\end{figure}

The full Hamiltonian in CP-decomposed form is then composed of $16$ nuclear potential terms (the CP rank of $V$ times two electrons), $8$ two-electron interaction terms, and $6$ kinetic energy operator terms corresponding to the three Cartesian directions per electron. The energy is minimized variationally with respect to the wavefunction CP factors using alternating least squares, analogously to the DMRG optimization but in a Cartesian-product rather than chain geometry. Figure~\ref{fig:placeholder} shows the convergence of the variational energy with respect to wavefunction rank $R_{\psi}$. The error is measured relative to the exact nonrelativistic ground-state energy of helium, $-2.903724$~Hartree, i.e.\ the complete-basis-set (CBS). It therefore includes both the discretization error of the grid ($\Delta x = 0.1$ Bohr, $91^3$ points) and the compression error of the CP ansatz. Chemical accuracy (1 kcal/mol $\approx 1.6$ millihartree) is obtained with only rank 24. This compares to an error of $30$\% more than chemical accuracy with a lattice spacing of $0.1$ Bohr using exact integrals and a rank-42 representation\cite{jerke2018two}. The memory required to perform their calculation was around $120$ GB. The largest calculation here used about 1.4 GB of RAM.

This is consistent with the work of Refs.~\citenum{huggins2025efficient, Kosugi2025}, which show that wavefunctions of chemical systems can be efficiently compressed using tensor representations. Explicitly, the rank-24 representation requires storing only $24 \times 3 \times 91^2 \approx 6\times 10^5$ elements ($\approx 4.8$~MB in double precision), compared to $91^6 \approx 6\times 10^{11}$ elements ($\approx 4.5$~TB) for the full statevector.

\subsection{CP decomposition of a pseudopotential}
Pseudopotentials are a widely used approximation in quantum chemistry that replace the explicit treatment of core electrons with an effective potential. In an all-electron calculation, every electron must be accounted for, but core electrons contribute little to chemical bonding and reactivity. Pseudopotentials capture the combined effect of the core electrons through a smooth, pre-fitted effective potential and treat only the valence electrons explicitly. This reduces the required basis set by eliminating the need to resolve the sharp wavefunction features near the nucleus. The result is a method that retains chemically accurate descriptions of bonding and reactivity at a fraction of the computational cost of a full all-electron treatment. As an example of how the CP decomposition can create efficient representations of pseudopotentials, we decompose the HGH\cite{hartwigsen1998relativistic} pseudopotential for Lithium. This local potential, in real space, is of the form 

\begin{eqnarray}\label{eq1}
V_{\rm loc}(r)&=&\frac{-Z_{\rm ion}}{r}
\mbox{erf}\left(\frac{r}{\sqrt{2}{r_{\rm loc}}}\right) 
+ \exp\left[-\frac{1}{2}\left(\frac{r}{{r_{\rm loc}}}\right)^2\right]
\nonumber\\
&&\times \left[ C_1 
             +C_2 \left(\frac{r}{{r_{\rm loc}}}\right)^2
             +C_3 \left(\frac{r}{{r_{\rm loc}}}\right)^4
             +C_4 \left(\frac{r}{{r_{\rm loc}}}\right)^6
       \right]
\end{eqnarray}
where $\mathrm{erf}$ denotes the error function and $Z_{\rm ion}$ is the ionic
charge of the atomic core, i.e., the total nuclear charge minus the number of
frozen core electrons.  The nonlocal part of the pseudopotential, $V_l({\bf
r},{\bf r}')$, is a sum of separable terms
\begin{equation}\label{eq2}
 V_l({\bf r},{\bf r}')= \sum_{i=1}^3 \sum_{j=1}^3
\sum_{m=-l}^{+l} Y_{l,m}(\hat{\bf r})\; p_i^l(r)\;  h_{i,j}^l\;
p_j^l(r')\;  Y^*_{l,m}(\hat{\bf r}')
\end{equation}
where $Y_{l,m}$ are the spherical harmonics and $l$ the angular
momentum quantum number with $h_{i,j}^{l}$ being the coefficients of the expansion.  The projectors $p_i^l(r)$ are Gaussians of
the form
\begin{equation}
p_i^l(r)=\frac{\sqrt{2} r^{l+2(i-1)}\exp\left(-\frac{r^2}{2{r_l}^2}\right)}
{{r_l}^{l+(4i-1)/2} \sqrt{\Gamma(l+\frac{4i-1}{2}) }}
\end{equation} 
where $\Gamma$ denotes the gamma function.  The projectors satisfy the
normalisation condition
\begin{equation}\label{eq5}
\int_0^\infty p_i^l(r) p_i^l(r)r^2dr=1 \; .
\end{equation}

The non-local contribution is directly expressible in CP format, decomposed by Cartesian component. The $l=0$ term is 
\begin{equation}
    V_0({\bf r},{\bf r}') = \frac{h^0_{11}}{\pi^{3/2}r_0^3}\prod_{\alpha,\alpha'\in\{x,y,z\}}\exp \left(-\frac{\alpha^2}{2r_0^2}\right)\exp \left(-\frac{\alpha'^2}{2r_0^2}\right)
\end{equation}
while the $l=1$ component is
\begin{equation}
    V_1({\bf r},{\bf r}') = \frac{2h^1_{11}}{\pi^{3/2}r_1^5}\left(x x'+y y'+z z'\right)\exp \left(-\frac{{\bf r}^2+{\bf r}'^2}{2r_1^2}\right)
\end{equation}
so the $l=0$ non-local contribution is rank-1 while the $l=1$ contribution is rank-3. These follow from Eq.~(\ref{eq2}) with $i=j=1$, using $Y_{00}=1/\sqrt{4\pi}$ and $\sum_m Y_{1m}(\hat{\bf r})Y^*_{1m}(\hat{\bf r}')=\frac{3}{4\pi}\hat{\bf r}\cdot\hat{\bf r}'$. Each is a product of a function of ${\bf r}$ and a function of ${\bf r}'$. Because $\exp[-({\bf r}^2+{\bf r}'^2)/2r_l^2]$ factorizes over Cartesian components, every term is a product over $x$, $y$, $z$ of one-dimensional matrices such as $xe^{-x^2/2r_1^2}\,\times x'\,e^{-x'^2/2r_1^2}$. For the Li HGH parameters used here, only the $i=j=1$ projectors are nonzero. Projectors with $i>1$ would contain $r^{2(i-1)}$ which are also products of 1D terms but would increase the rank further.

\begin{figure}
    \centering
    \includegraphics[width=0.9\linewidth]{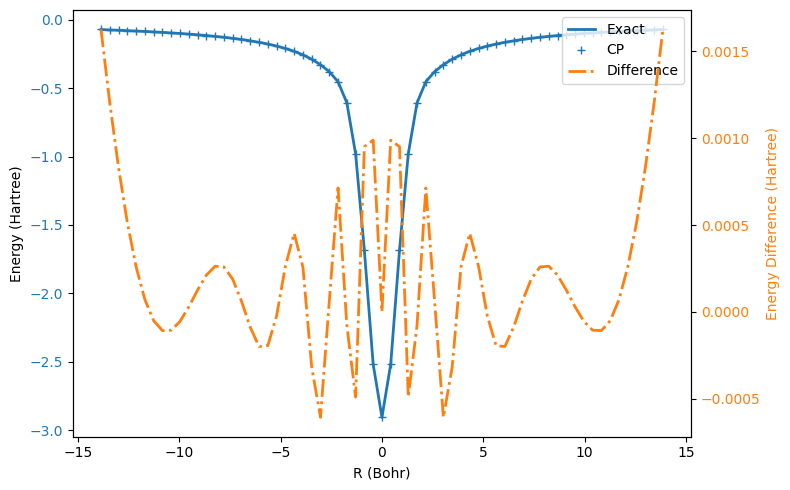}
    \caption{The error of a rank-4 CP representation (with respect to Cartesian coordinates) of the Li HGH pseudopotential with grid spacing $0.25$ Bohr along the (1,1,1) direction.}
    \label{fig:cp_for_li}
\end{figure}

The local term is not obviously separable but we can perform the same decomposition as was done for the sinc-derived Coulomb potential at grid points. We can see in Figure \ref{fig:cp_for_li} that the error for a rank-4 decomposition with $\Delta x=0.25$ Bohr is low along the whole (1,1,1) direction and only relatively large at the corners. Unfortunately, the decomposition will need to be performed separately for each grid-spacing and pseudopotential due to the error function scaling not being separable. The expansion being low-rank is not surprising given the known decomposition using B{\"u}rmann's theorem\cite{schopf2014burmann}.

It should be noted that there has been previous work\cite{Shokrian_Zini_2023, Berry_2024} studying the applicability of pseudopotentials to reduce the cost of quantum simulation of materials in first-quantization but a constant-factor comparison will be left to future work.

\section{\label{sec:lcu}Application to Quantum Computing with the Linear Combination of Unitaries Method}

The CP decomposition can be applied to quantum computing using the Linear Combination of Unitaries (LCU) framework, which provides a systematic approach to block-encoding general operators via two circuits, $PREP_O$ and $SELECT_O$.

The $PREP_O$ circuit performs:
\begin{equation}
    PREP_O\left|\psi\right\rangle \left|0\right\rangle^a = \frac{1}{\beta}\sum_j \left|\psi\right\rangle \sqrt{\left|a_j\right|}\left|j\right\rangle,
\end{equation}
where the operator is $O=\sum_{j=0}^{N_o-1} a_j U_j$ with all $a_j$ positive real constants (signs absorbed into unitaries $U_j$), and $\beta=\sqrt{\sum_j a_j}$. Here $N_o$ is the number of terms in the operator.

The $SELECT_O$ operation is:
\begin{equation}
    SELECT_O\frac{1}{\beta}\sum_j \left|\psi\right\rangle \sqrt{\left|a_j\right|}\left|j\right\rangle =\sum_j \frac{1}{\beta}U_j\left|\psi\right\rangle \sqrt{\left|a_j\right|}\left|j\right\rangle.
\end{equation}

The complete circuit $PREP_O \, SELECT_O \, PREP_O^{\dagger}$ implements the operator $O$ in the upper-left block of the matrix where all $a$ ancilla qubits are $\left|0\right\rangle$. This block encoding is then usable in quantum phase estimation (QPE) to extract eigenvalues of $O$, with the normalization factor $\beta$ determining the energy resolution achievable per QPE shot. The primary bottleneck is typically the $PREP_O$ operation, which can be performed with T-gates scaling linearly in the number of coefficients using the alias sampling method\cite{Babbush2018}.

Focusing on the kinetic energy term, $SELECT_T$ and $PREP_T$, the scaling is $\mathcal{O}(\eta \log^2 N)$ due to the controlled QFT required\cite{Su2021}. For diagonal potential terms in Eq.~(\ref{eq.regularized}), there are generally $N^2$ unitaries for the two-body interaction, yielding $\mathcal{O}(N^2)$ scaling for naive $PREP_V$. Reference~\citenum{Su2021} achieved $\mathcal{O}(N^{2/3})$ by partitioning the potential into equal-magnitude pieces and applying phases on-the-fly.

\subsection{Block Encoding the CP-Decomposed Potential}

\begin{figure}[htbp]
    \centering
    \begin{quantikz}
\lstick{\ket{0}$^3$}   & \gate{S_{prep}} & \ctrl{1}            & \ctrl{2}         & \ctrl{3}         &\rstick{\ket{s}}\\
\lstick{\ket{0}$^m$}   & \qw & \gate{C_{prep}}     & \qw              & \qw              &\rstick{\ket{c}}\\
\lstick{\ket{0}$^m$}   & \qw & \qw                 & \gate{C_{prep}}  & \qw              &\rstick{\ket{c}}\\
\lstick{\ket{0}$^m$}   & \qw & \qw                 & \qw              & \gate{C_{prep}}      &\rstick{\ket{c}}\\
\end{quantikz}
    \caption{The $PREP_V$ circuit for the CP-decomposed Coulomb potential.}
    \label{fig:uprep}
\end{figure}
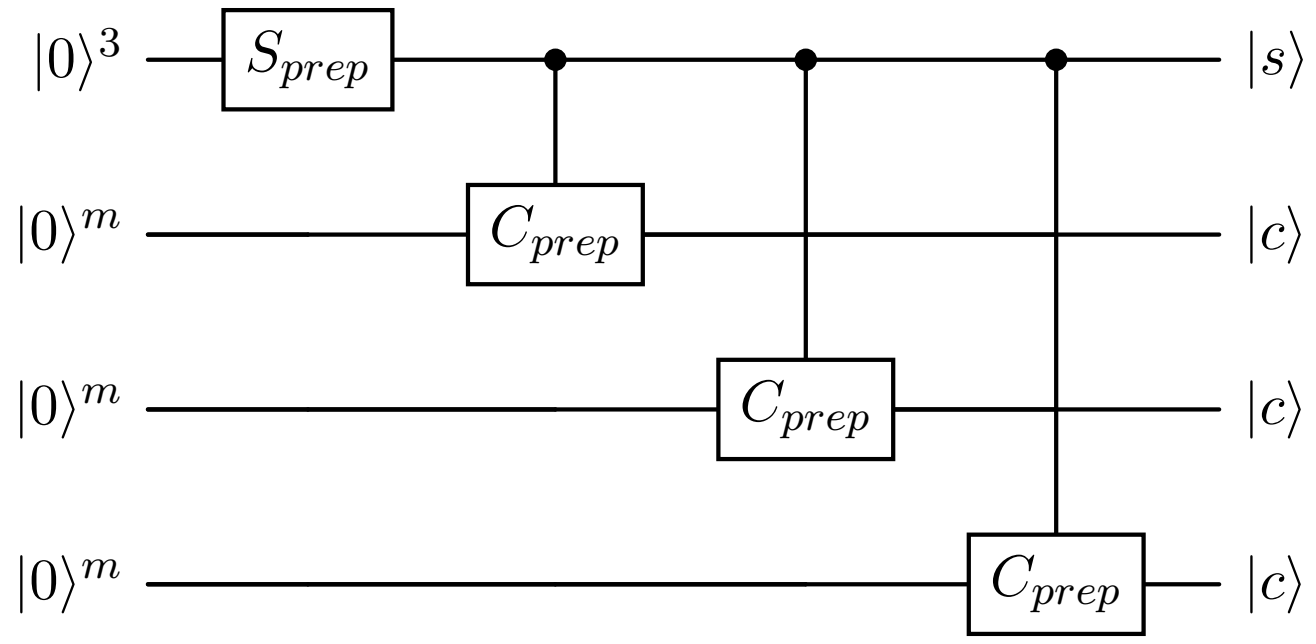

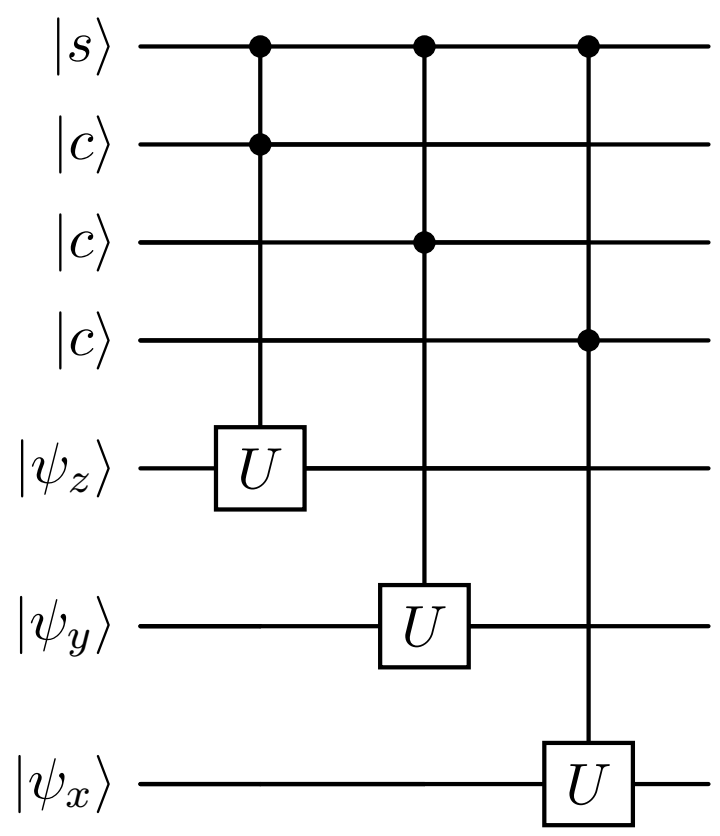
\begin{figure}[htbp]
    \centering
    \begin{quantikz}
\lstick{\ket{s}}   & \ctrl{1} & \ctrl{2} & \ctrl{3} &\\
\lstick{\ket{c}}   & \ctrl{3} & \qw      & \qw      &\\
\lstick{\ket{c}}   & \qw      & \ctrl{3} & \qw      &\\
\lstick{\ket{c}}   & \qw      & \qw      & \ctrl{3} &\\
\lstick{\ket{\psi_z}}   & \gate{U} & \qw      & \qw      &\\
\lstick{\ket{\psi_y}}   & \qw      & \gate{U} & \qw      &\\
\lstick{\ket{\psi_x}}   & \qw      & \qw      & \gate{U} & \\
\end{quantikz}
    \caption{The $SELECT_V$ operation for the CP-decomposed Coulomb potential. The registers $\ket{\psi_z}$, $\ket{\psi_y}$, $\ket{\psi_x}$ hold the $z$, $y$, $x$ components of the electronic position (one-body) or position-difference (two-body) index. For one-body terms, each $U$ applies the diagonal phase $c_{p}^{(r)}$ to the corresponding electronic register; for two-body terms, the registers are replaced by the pair $\ket{i_{x_1}}\ket{i_{x_2}}$ after quantum subtraction, and $U$ applies $c_{i_{x_1}-i_{x_2}}^{(r)}$.}
    \label{fig:uselect}
\end{figure}

The one-electron CP-decomposed Coulomb potential can be block-encoded using the $PREP_V$ and $SELECT_V$ circuits shown in Figs.~\ref{fig:uprep} and~\ref{fig:uselect}. These can be derived from Ref \cite{kamakari2025higherordertensorfactorizations} by performing the three Cartesian preparations in parallel instead of sequentially, as noted as possible in that work. The parallelism is valid because the $x$, $y$, and $z$ factor vectors are prepared independently into separate ancilla registers that are never entangled with one another during $PREP_V$. Each coefficient register requires $m=\log_2(N^{1/3})$ qubits for one-body terms; for two-body terms the index runs over differences $i_p - i_q \in [-N^{1/3}, N^{1/3}]$, requiring one additional qubit per register.

The $S_{prep}$ circuit prepares a 3-qubit register (for the 8 rank terms):
\begin{equation}
    S_{prep}\ket{0}^3 = \frac{1}{A}\sum_{r=0}^7\sqrt{s_r \|c_r\otimes c_r \otimes c_r\|_1}\ket{r}=\ket{s}
\end{equation}
where $A = \sqrt{\sum_r s_r \|c_r\|_1^3}$ is the normalization constant and the $\ell_1$ norm appears because the alias sampling method prepares amplitudes proportional to $\sqrt{a_j}$ so that the block-encoding normalization is $\beta = \sum_r s_r \|c_r\|_1^3$. The $C_{prep}$ circuit then prepares each Cartesian coefficient register conditioned on $\ket{r}$:
\begin{equation}
    C_{prep}\ket{r}\ket{0}^m = \ket{r}\frac{1}{\|c_r\|_1^{1/2}}\sum_{k}\sqrt{|c_r^{(k)}|}\ket{k},
\end{equation}
with the sign of $c_r^{(k)}$ absorbed into the $SELECT_V$ unitary.

For the two-body terms, each term depends only on the difference between the two position registers. Therefore, one applies a quantum subtraction operation\cite{Su2021} $S\ket{i}\ket{j}\rightarrow \ket{i}\ket{i-j}$, selects on $i-j$ via the coefficient vector $c_{i-j}$, and then uncomputes the subtraction to restore $\ket{i}\ket{j}$. This avoids the need for separate alias sampling tables for each pair of grid indices.

Further optimizations are possible for $SELECT_V$. The decomposition of each $c$ vector as a diagonal qubit operator is generally dense, involving all products of $Z$ and $I$ Pauli operators. The only adjustment required across rank terms is a sign change conditioned on $\ket{s}$. The extra controls on $\ket{s}$ for each $U$ can be eliminated by working with unsigned Pauli products and applying controlled phase gates to only one of the three $\ket{c}$ registers, since all three Cartesian components are equivalent.

\subsection{T-Gate Scaling of single-electron potential}

We estimate the T-gate cost of the CP-decomposed block encoding using Qualtran\cite{harrigan2024qualtran} for various values of $N^{1/3}$. The $PREP$ operation uses the alias sampling method\cite{Babbush2018} for three controlled applications. The \emph{GenericSelect} method implements the $SELECT$ operation.

\begin{figure}[htbp]
    \centering
    \includegraphics[width=0.7\textwidth]{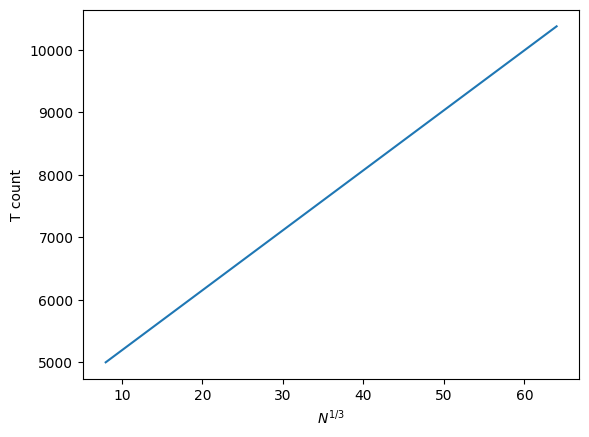}
    \caption{T-gate count for a single application of a one-body Coulomb term using CP decomposition for different values of $N^{1/3}$, demonstrating $\mathcal{O}(N^{1/3})$ scaling.}
    \label{fig:tgate_scaling}
\end{figure}

Figure~\ref{fig:tgate_scaling} shows the $\mathcal{O}(N^{1/3})$ T-gate scaling for a single one-body Coulomb application calculated using Qualtran\cite{harrigan2024qualtran}. For the full first-quantized Hamiltonian with $\eta$ electrons and $N_A$ nuclei, the nuclear-electron interaction contributes $\eta N_A$ such applications per Hamiltonian walk step, so the dominant cost scales as $\mathcal{O}(\eta N_A N^{1/3})$. The cost is dominated by the linear scaling alias sampling of the length-$N^{1/3}$ coefficient vectors. The quantum Fourier transforms and adders in the circuit act on registers of only $\log_2 N^{1/3}$ qubits with their cost scaling with the register width.  The electron-electron interaction has the same asymptotic per-term cost by generating the transformation $\ket{i}\ket{j}\rightarrow \ket{i}\ket{i-j}$ and applying the larger $(2N^{1/3}-1)$-length operations for each of $(x_1-x_2),\, (y_1-y_2),\ (z_1-z_2)$ and then unitarily transforming back.

\section{\label{sec:geometry}Application to Molecular Geometry Sampling}

Having established an efficient block encoding of the Coulomb potential, we now discuss its application to sampling molecular geometries. In Ref.~\citenum{kosugi2023exhaustive}, probabilistic imaginary-time evolution in a quasi-non-Born-Oppenheimer setting was proposed as a way to probe molecular geometries, with the treatment of the Coulomb potential left to future work. The present work addresses that gap. Ref.~\citenum{kosugi2023exhaustive} (and follow-up work\cite{Nishiya_2024}) used Trotterization to implement imaginary-time steps; here we instead use the block-encoding framework developed above, which enables the use of the more rapidly converging Coulomb-like potentials. Rather than imaginary-time evolution, we employ the low-energy-state preparation algorithm of Lin and Tong\cite{LinTong2020} within the quantum signal processing framework\cite{Dong_2021}.

Treating the nuclei as degrees of freedom (although with no kinetic energy contribution) connects this approach to earlier work on simulating chemistry without the Born-Oppenheimer approximation on quantum computers. Kassal \emph{et al.}\cite{Kassal2008} showed that grid-based, first-quantized simulation of electrons and nuclei together scales polynomially and can be more efficient than a Born-Oppenheimer treatment for systems with more than a few atoms. More recently, Eklund \emph{et al.}\cite{eklund2026endtoendsimulationchemicaldynamics} developed an end-to-end algorithm for non-adiabatic chemical dynamics with bounded error, with complexity sublinear in the grid size. The geometry-sampling scheme below can be viewed as a static (ground-state-filtered) version of that non-Born-Oppenheimer picture, with nuclear kinetic energy dropped.

A key point is that the improved convergence using specialized Coulomb potentials enables exploration of configuration space with sufficient resolution to identify energetically favorable geometries without necessarily being quantitatively accurate. Classically, the standard approach to molecular geometry optimization is to perform a sequence of energy evaluations at different nuclear configurations, guiding the search via gradients or heuristics. On a quantum computer, one can instead encode a superposition of nuclear configurations into a quantum register and use the Hamiltonian evolution or ground-state preparation to bias the measurement outcome toward low-energy geometries.\cite{kosugi2023exhaustive} This quantum-parallel approach is particularly appealing for rugged potential energy surfaces where classical gradient methods become trapped in local minima.

\subsection{Geometry Sampling in Momentum Space}

Working in momentum space for the electronic degrees of freedom is advantageous for encoding nuclear position dependence, because the Coulomb potential in momentum space takes a multiplicative form modulated by a phase factor that depends linearly on nuclear coordinates. This linear dependence makes it straightforward to place nuclear positions in quantum superposition and apply the corresponding phase kick to the electronic momentum register, as we now describe. The matrix elements are
\begin{equation}
\begin{array}{rl}
    T = & \sum_{p\in G}\frac{n_p^2}{2}\ket{p}\bra{p}_i, \\
    V = & -\sum_{i=1}^{\eta}\sum_{l=1}^{N_A}\zeta_l \sum_{p\in G, p^{\prime} \in G}\bar{V}_{p-p^{\prime}}e^{iR_l(p-p^{\prime})}\ket{p^{\prime}}\bra{p}_i \\
    U = & \frac{1}{2}\sum_{i\neq j=1}^{\eta}\sum_{p,p',q,q'\in G}\bar{V}_{p'-p}\,\delta_{p'-p,\,q-q'}\ket{p'}\bra{p}_i\ket{q'}\bra{q}_j
\end{array}
\end{equation}
where $G$ represents the range of the indices but this time over momentum space.

It is important to note how the momentum-space matrix elements $\bar V_{p-p'}$ are obtained. We do not attempt to directly calculate (by integration) the analytic momentum-space Coulomb potential $4\pi/|\mathbf k|^2$ matrix elements. Instead, the potential is constructed and CP-decomposed in real space (Eq.~(\ref{eq:cp_decomp})), and $\bar V$ is then obtained by a discrete Fourier transform of the real-space potential,
\begin{equation}\label{eq:vbar_ft}
    \bar V_{p-p'} = \frac{1}{N}\sum_{\mathbf x\in G} V(\mathbf x)\,e^{-\iu(\mathbf k_p-\mathbf k_{p'})\cdot\mathbf x}
    = \frac{1}{N\Delta x}\sum_{r=1}^{8} s_r\, \tilde c^{(r)}_{n_x}\tilde c^{(r)}_{n_y}\tilde c^{(r)}_{n_z},
\end{equation}
where $\tilde c^{(r)}$ is the one-dimensional discrete Fourier transform of $c^{(r)}$ and $(n_x,n_y,n_z)$ are the Cartesian components of $p-p'$. The Fourier transform of a product of one-dimensional fucntions transforms as a product of one-dimensional functions so the CP rank is unchanged by this step.  We find numerically that this route gives better results then direct integration. Using direct integration produces results similar to using the soft-core potential for the Hydrogen system studied in Table \ref{tab:convergence}.

Following the exhaustive search approach, we add a superposition over nuclear coordinates $l$ without including kinetic energy for these terms:
\begin{equation}
\begin{array}{rl}
    T = & \sum_{p\in G}\frac{n_p^2}{2}\ket{p}\bra{p}_i \\
    V = & -\sum_{i=1}^{\eta}\sum_{l=1}^{N_A}\zeta_l \sum_{\substack{p,p^{\prime} \in G \\ q\in J}}\bar{V}_{p-p^{\prime}}e^{i(R_qq-R_{l}q^{\prime})p^{\prime}}\ket{q^{\prime}}_l\ket{p^{\prime}}_i\bra{p}_i\bra{q}_l \\
    U = & \frac{1}{2}\sum_{i\neq j=1}^{\eta}\sum_{p,p',q,q'\in G}\bar{V}_{p'-p}\,\delta_{p'-p,\,q-q'}\ket{p'}\bra{p}_i\ket{q'}\bra{q}_j  \\
    U_{nuc} = & \frac{1}{2}\sum_{l\neq l^{\prime}=1}^{N_A}\zeta_l\zeta_{l^{\prime}}\sum_{p,q\in J}V_{p,q}\ket{p}_l\bra{p}_l\ket{q}_{l^{\prime}}\bra{q}_{l^{\prime}}
\end{array}
\end{equation}
where $J$ indexes the nuclear coordinate superposition.  In momentum space, the electron-electron interaction is not diagonal as it transfers momentum $p'-p$ from electron $j$ to electron $i$, with matrix elements that depend only on the momentum difference, so that $U$ is Toeplitz in each momentum register. The nuclear registers, in contrast, hold positions, so $U_{nuc}$ is diagonal and uses the real-space kernel $V_{p,q}$ evaluated at the separation of the nuclear positions $R_p$ and $R_q$.

Focusing on the nuclear-electronic interaction $V$, we can rewrite it using standard matrix-vector techniques:
\begin{equation}
    V =  -\sum_{i=1}^{\eta}\sum_{l=1}^{N_A}\zeta_l \sum_{p^{\prime},q^{\prime}}e^{-iR_lq^{\prime}p^{\prime}} \ket{q^{\prime}}_l\sum_{p,p^{\prime} \in G}\bar{V}_{p-p^{\prime}}\ket{p^{\prime}}_i \sum_{p,q}e^{iR_lqp}\bra{p}_i\bra{q}_l
\end{equation}

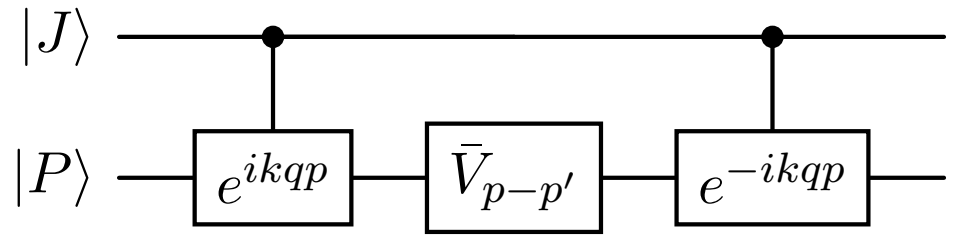
\begin{figure}[htbp]
    \centering
    \begin{quantikz}
\lstick{\ket{J}} & \ctrl{1} & \qw  & \ctrl{1}     & \\
\lstick{\ket{P}}&\gate{e^{ikqp}}      & \gate{\bar{V}_{p-p^{\prime}}} &\gate{e^{-ikqp}} &
\end{quantikz}
    \caption{Application of the nuclear potential for a nuclear register $\ket{J}$ and electronic register $\ket{P}$.}
    \label{fig:nucleus_electron}
\end{figure}

This structure enables applying a superposition of nuclear coordinates in real space to momentum-space electrons, as illustrated in Fig.~\ref{fig:nucleus_electron}.

The phase application circuit is constructed using standard phase estimation techniques:

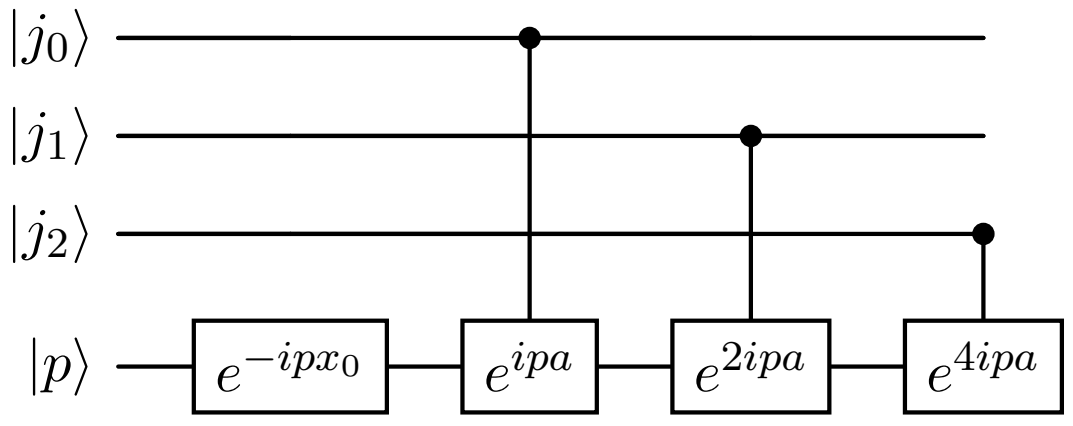
\begin{figure}[htbp]
    \centering
    \begin{quantikz}          
\lstick{\ket{j_0}}   & \qw & \ctrl{3}     & \qw              & \qw              \\
\lstick{\ket{j_1}}   & \qw & \qw                 & \ctrl{2}  & \qw              \\
\lstick{\ket{j_2}}   & \qw & \qw                 & \qw              & \ctrl{1}      \\
\lstick{\ket{p}}   & \gate{e^{-ipx_0}} & \gate{e^{ipa}}     & \gate{e^{2ipa}}              & \gate{e^{4ipa}}              \\
\end{quantikz}
    \caption{Circuit to generate phases for the momentum-space potential, analogous to the standard phase-building circuit in phase estimation for a linear potential $x$.}
    \label{fig:phase_circuit}
\end{figure}

\subsection{Small Example}
To provide a small example of how this may work in practice, we apply the ground-state preparation function directly to a single particle in a 1D pseudo-Coulomb system, using the matrix representation of the Hamiltonian. We do not explicitly construct the quantum circuits here, but the tools developed above make a full constant-factor circuit analysis possible in future work. The nuclear potential is in atomic units,

\begin{equation}
    V(x-a)=\frac{4}{10}\exp\left[-(x-a)^2\right],
\end{equation}
and the total Hamiltonian is
\begin{equation}
    H_p=V(x-a)+V(x-b)+V(a-b)
\end{equation}
where $V(a-b)$ represents the nuclear energy (which will be treated classically) with matrix elements,
\begin{equation}
    V_{k,l} = \frac{4}{10}\exp\left[-((k-l)\mathrm{d}x)^2\right]
\end{equation}

The particle is treated in momentum space, the potential is $V(p,a)=e^{ipa}e^{-p^2/4}$  with grid spacing which for the grid-based Hamiltonian here has matrix elements,
\begin{equation}
    \bar{V}_{i,j}(p;a) = -\mathrm{d}p \frac{2}{10}\sqrt{\frac{1}{\pi}}\exp\!\left[-\iu \mathrm{d}p(i-j)a - \left(\mathrm{d}p(i-j)\right)^2/4\right]
\end{equation}
where $\mathrm{d}p=2\pi/(dx \cdot N_1)$ with $dx=1$ Bohr and $N_1=64$ is the momentum grid spacing, and kinetic energy matrix elements,
\begin{equation}
    T_{i,j} = \delta_{i,j} \left(\mathrm{d}p\cdot i - N_1/2\right)^2.
\end{equation}

We utilize the fixed-point ground state preparation function based on Chebyshev polynomials\cite{LinTong2020}:
\begin{equation}
    f(H,\Delta,K)=\sum_{n=0}^{K}\frac{T_n\!\left(-1+2\frac{H^2-\Delta^2}{1-\Delta^2}\right)}{T_n\!\left(-1+2\frac{-\Delta^2}{1-\Delta^2}\right)}
\end{equation}
where $T_n$ denotes the Chebyshev polynomial of the first kind of degree $n$, $\Delta$ is a gap parameter that controls the spectral filter sharpness, and $K$ is the polynomial degree (which sets the circuit depth). The function $f$ approximately projects onto the ground state subspace by suppressing eigenvalues with $|E| > \Delta$ while amplifying those with $|E| \ll \Delta$, assuming the spectrum is normalized to $[-1,1]$. We use $\Delta=0.1$ and apply it directly to the Hamiltonian with matrix elements
\begin{equation}
    H_{i,j,a,b}=T_{i,j}+\delta_{a}\bar{V}_{i,j}(p;a)+\delta_{b}\bar{V}_{i,j}(p; b)+V_{a,b}.
\end{equation}
followed by normalization to ensure its spectrum lies in $[-1,1]$.

\begin{figure}[htbp]
    \centering
    \includegraphics[width=0.9\linewidth]{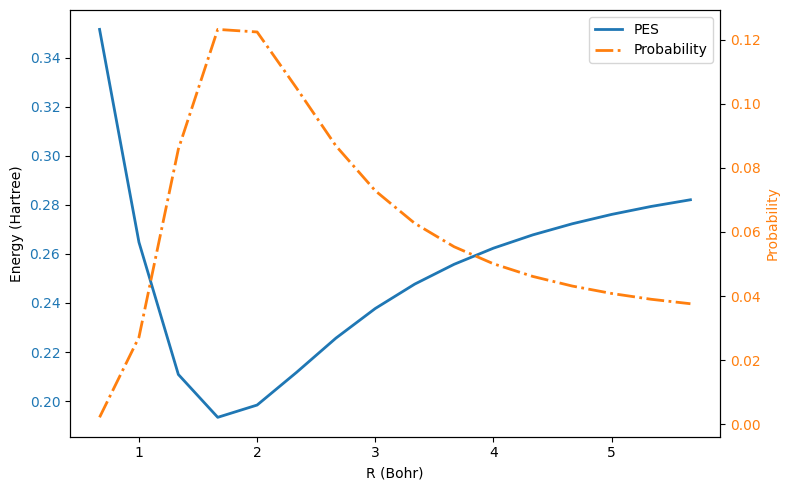}
    \caption{Probability distribution over nuclear coordinate configurations after applying the Chebyshev ground-state preparation filter $f(H,0.1,60)$ to a uniform initial state. The pseudo-Coulomb potential $V(x-a)+V(x-b)$ with $R=b-a$ for $a=-\frac{1}{3}$ Bohr and $b=\frac{j}{3}$ Bohr for $j\in[1,16]$. The enhanced probability near the energy minimum demonstrates that measuring the nuclear register preferentially yields low-energy configurations.}
    \label{fig:sampling_probability}
\end{figure}

The matrix $F(H, 0.1)$ is then applied to the initial state, which is taken to be uniform over real-space nuclear coordinates and with the electronic degree of freedom initialized to zero momentum ($s_e = \delta_{k,0}$). After applying $F$ and normalizing, the squared amplitudes over nuclear coordinate configurations give a probability distribution that is peaked near the minimum-energy geometry. The result with $K=60$ is shown in Fig.~\ref{fig:sampling_probability}.

As expected, the probability of measuring lower-energy nuclear configurations is significantly higher than for higher-energy ones, confirming that the Chebyshev filter successfully biases the distribution toward the ground-state geometry.

\subsection{Enabling Geometry Exploration}

The efficient CP decomposition of the Coulomb potential dramatically reduces the computational cost per basis function. This reduction has several important consequences for molecular geometry sampling:

\begin{enumerate}
    \item \textbf{Reduced Basis Set Requirements}: The $\mathcal{O}(N^{1/3})$ scaling with simplified circuits means that moderately sized molecules can be simulated with manageable quantum resources.
    
    \item \textbf{Superposition Over Configurations}: By encoding nuclear positions in quantum registers (as in the momentum-space formulation), we can explore multiple geometries simultaneously through quantum parallelism.
    
    \item \textbf{Amplitude Amplification}: The reduced cost per evaluation makes fixed-point amplitude amplification practical for boosting the success probability of finding low-energy configurations\cite{brassard2002quantum}.
\end{enumerate}

The combination of efficient Coulomb potential evaluation and quantum parallelism over configuration space provides a pathway to addressing the exponential scaling challenge of molecular geometry optimization on quantum computers.

\section{Application to Time Evolution\label{sec:timeevolution}}

Time evolution in chemistry is one of the most promising candidates for quantum computers and, although not the focus of this manuscript, is important to discuss. Here we estimate how a faster-converging Coulomb representation can help with this instance as well.

With qubitization\cite{LowChuang2019}, the evolution $e^{-\iu Ht}$ can be implemented to error $\epsilon$ with
\begin{equation}
    Q(t,\epsilon) = \mathcal{O}\!\left(\lambda t + \log(1/\epsilon)\right)
\end{equation}
queries to the block encoding where $\lambda$ is the normalization. Here $\lambda = \lambda_T + \lambda_V + \lambda_U$ and the T-gate cost per unit of time can be approximated as $\mathcal{O}\left(\lambda C_{\mathrm{BE}}\right)$
with $C_{\mathrm{BE}}$ the T-cost of one block-encoding. For a simulation box of fixed side length $L$ with $N = (L/\Delta x)^3$ grid points per particle. The dominant contribution to the normalization is the kinetic which scales as $\lambda_T \propto \eta/\Delta x^2$. The Coulomb contribution scales as $\eta N_A N^{1/3} \propto \eta N_A L/\Delta x$ (Sec.~\ref{sec:lcu}). As an example, assume the improved potential reaches the target accuracy at a grid spacing that is larger by a factor $\alpha$. Table~\ref{tab:convergence} suggests that $\alpha \approx 2$ is not unreasonable. Then, up to lower-order terms, the T-gates fall by roughly $\alpha^2$ (from $\lambda_T$) times $\alpha$ (from $C_{\mathrm{BE}}$), that is, $\sim\alpha^3 \approx 8$ for $\alpha=2$. The logical qubit count per particle also falls by approximately $3$ per particle.
For quantum phase-estimation, the frequency resolution is $\delta\omega \sim 2\pi/t_{\max}$. The number of queries needed for energy precision $\varepsilon$ is $\sim\pi\lambda/(2\varepsilon)$\cite{Babbush2018}. As above, the accuracy will increase as $\sim\alpha^3$. 

These estimates neglect the $\lambda_V$ and $\lambda_U$ contributions and do not include the constant factor analysis.  They simply show that the gains in accuracy enter the cost of time evolution to roughly the third power. A natural target for these basis functions is non-Born-Oppenheimer dynamics of electrons and nuclei on a common grid\cite{Kassal2008, eklund2026endtoendsimulationchemicaldynamics}. For a more concrete estimate, the potential of Eq.~(\ref{eq:vbar_ft}) could be included in the constant factor estimate of Ref.~\citenum{eklund2026endtoendsimulationchemicaldynamics}.

\section{\label{sec:conclusion}Conclusion}

We have demonstrated that the Coulomb potential in real space can be efficiently decomposed into a canonical polyadic tensor with low-rank Cartesian components. This decomposition enables relatively simple block-encoding circuits with $\mathcal{O}(N^{1/3})$ T-gate scaling, representing a substantial improvement over naive implementations. The method generalizes to Coulomb-type potentials broadly and can be applied to various potential forms, many of which have superior convergence properties compared to standard regularization techniques\cite{Lehtola2023}.

Crucially, this efficiency gain makes practical the application of quantum algorithms to probe molecular geometries for moderately sized molecules. The reduced computational cost per basis function, combined with the ability to encode nuclear positions in quantum superposition, provides a viable approach to the exponentially scaling problem of molecular configuration space exploration.

Beyond geometry sampling, another long-term benefit is likely in time evolution. Our estimate in Sec.~\ref{sec:timeevolution} indicates that the T-gate cost per unit of simulated time all improve roughly as the cube of the gain in usable grid spacing.

Future work may further reduce T-gate costs for two-body terms by exploiting the Toeplitz structure of the matrices\cite{Wan2021, Shang2023}. This would involve transforming the register $\ket{i_{x_1}}\ket{i_{x_2}}\ket{0}$ to $\ket{i_{x_1}}\ket{i_{x_2}-i_{x_1}}$, applying the Toeplitz matrix, and reversing the transformation. Additionally, integration with quantum optimization algorithms and amplitude amplification techniques promises to enhance the practical utility of these methods for computational chemistry applications.

\section{Acknowledgements}

JDW is supported by the Department of Energy under award DE-SC0019374 and by the Army Research Office under award W911NF2410043.

\bibliography{references}

@article{hartwigsen1998relativistic,
  title = {Relativistic separable dual-space Gaussian pseudopotentials from {{H}} to {{Rn}}},
  author = {Hartwigsen, C. and Goedecker, S. and Hutter, J.},
  journal = {Phys. Rev. B},
  volume = {58},
  issue = {7},
  pages = {3641--3662},
  numpages = {0},
  year = {1998},
  month = {Aug},
  publisher = {American Physical Society},
  doi = {10.1103/PhysRevB.58.3641},
}

@misc{harrigan2024qualtran,
    title={Expressing and Analyzing Quantum Algorithms with Qualtran},
    author={Matthew P. Harrigan and Tanuj Khattar
        and Charles Yuan and Anurudh Peduri and Noureldin Yosri
        and Fionn D. Malone and Ryan Babbush and Nicholas C. Rubin},
    year={2024},
    eprint={2409.04643},
    archivePrefix={arXiv},
    primaryClass={quant-ph},
    doi={10.48550/arXiv.2409.04643},
    url={https://arxiv.org/abs/2409.04643},
}

@misc{kamakari2025higherordertensorfactorizations,
      title={Higher order tensor factorizations for block encoding vibrational and vibronic Hamiltonians}, 
      author={Hirsh Kamakari and Emil Zak},
      year={2025},
      eprint={2504.08065},
      archivePrefix={arXiv},
      primaryClass={quant-ph},
      url={https://arxiv.org/abs/2504.08065}, 
}

@article{jerke2018two,
    author = {Jerke, Jonathan and Poirier, Bill},
    title = {Two-body Schrödinger wave functions in a plane-wave basis via separation of dimensions},
    journal = {The Journal of Chemical Physics},
    volume = {148},
    number = {10},
    pages = {104101},
    year = {2018},
    month = {03},
    issn = {0021-9606},
    doi = {10.1063/1.5017621},
}

@article{huggins2025efficient,
  title = {Efficient State Preparation for the Quantum Simulation of Molecules in First Quantization},
  author = {Huggins, William J. and Leimkuhler, Oskar and Stetina, Torin F. and Whaley, K. Birgitta},
  journal = {PRX Quantum},
  volume = {6},
  issue = {2},
  pages = {020319},
  numpages = {54},
  year = {2025},
  month = {Apr},
  publisher = {American Physical Society},
  doi = {10.1103/PRXQuantum.6.020319}
}

@article{Shokrian_Zini_2023,
   title={Quantum simulation of battery materials using ionic pseudopotentials},
   volume={7},
   ISSN={2521-327X},
   DOI={10.22331/q-2023-07-10-1049},
   journal={Quantum},
   publisher={Verein zur Forderung des Open Access Publizierens in den Quantenwissenschaften},
   author={Shokrian Zini, Modjtaba and Delgado, Alain and dos Reis, Roberto and Moreno Casares, Pablo Antonio and Mueller, Jonathan E. and Voigt, Arne-Christian and Arrazola, Juan Miguel},
   year={2023},
   month={7}, pages={1049} }

@article{Berry_2024,
   title={Quantum simulation of realistic materials in first quantization using non-local pseudopotentials},
   volume={10},
   ISSN={2056-6387},
   DOI={10.1038/s41534-024-00896-9},
   number={1},
   journal={npj Quantum Information},
   publisher={Springer Science and Business Media LLC},
   author={Berry, Dominic W. and Rubin, Nicholas C. and Elnabawy, Ahmed O. and Ahlers, Gabriele and DePrince, A. Eugene and Lee, Joonho and Gogolin, Christian and Babbush, Ryan},
   year={2024},
   month={12} }

@article{Dong_2021,
  title = {Efficient phase-factor evaluation in quantum signal processing},
  author = {Dong, Yulong and Meng, Xiang and Whaley, K. Birgitta and Lin, Lin},
  journal = {Phys. Rev. A},
  volume = {103},
  issue = {4},
  pages = {042419},
  numpages = {22},
  year = {2021},
  month = {Apr},
  publisher = {American Physical Society},
  doi = {10.1103/PhysRevA.103.042419},
}

@article{Schollwock2011,
   title={The density-matrix renormalization group in the age of matrix product states},
   volume={326},
   ISSN={0003-4916},
   url={http://dx.doi.org/10.1016/j.aop.2010.09.012},
   DOI={10.1016/j.aop.2010.09.012},
   number={1},
   journal={Annals of Physics},
   publisher={Elsevier BV},
   author={Schollw{\"o}ck, Ulrich},
   year={2011},
   month={1}, pages={96-192} }

@article{Oseledets2011,
author = {Oseledets, I. V.},
title = {Tensor-Train Decomposition},
journal = {SIAM Journal on Scientific Computing},
volume = {33},
number = {5},
pages = {2295-2317},
year = {2011},
doi = {10.1137/090752286}
}

@misc{brassard2002quantum,
   title={Quantum amplitude amplification and estimation},
   ISBN={9780821878958},
   ISSN={0271-4132},
   DOI={10.1090/conm/305/05215},
   journal={Quantum Computation and Information},
   publisher={American Mathematical Society},
   author={Brassard, Gilles and Høyer, Peter and Mosca, Michele and Tapp, Alain},
   year={2002},
   pages={53-74} }

@article{Kosugi2025,
  title = {Tensor-decomposition technique for qubit encoding of maximal-fidelity Lorentzian orbitals in real-space quantum chemistry},
  author = {Kosugi, Taichi and Huang, Xinchi and Nishi, Hirofumi and Matsushita, Yu-ichiro},
  journal = {Phys. Rev. A},
  volume = {111},
  issue = {5},
  pages = {052615},
  numpages = {19},
  year = {2025},
  month = {May},
  publisher = {American Physical Society},
  doi = {10.1103/PhysRevA.111.052615},
  url = {https://link.aps.org/doi/10.1103/PhysRevA.111.052615}
}

@article{kosugi2023exhaustive,
   title={Exhaustive search for optimal molecular geometries using imaginary-time evolution on a quantum computer},
   volume={9},
   ISSN={2056-6387},
   DOI={10.1038/s41534-023-00778-6},
   number={1},
   journal={npj Quantum Information},
   publisher={Springer Science and Business Media LLC},
   author={Kosugi, Taichi and Nishi, Hirofumi and Matsushita, Yu-ichiro},
   year={2023},
   month={11} }

@article{Beylkin2002,
author = {Gregory Beylkin  and Martin J. Mohlenkamp },
title = {Numerical operator calculus in higher dimensions},
journal = {Proceedings of the National Academy of Sciences},
volume = {99},
number = {16},
pages = {10246-10251},
year = {2002},
doi = {10.1073/pnas.112329799}}

@article{Babbush2018,
   title={Encoding Electronic Spectra in Quantum Circuits with Linear {{T}} Complexity},
   volume={8},
   ISSN={2160-3308},
   doi={10.1103/physrevx.8.041015},
   number={4},
   journal={Physical Review X},
   publisher={American Physical Society (APS)},
   author={Babbush, Ryan and Gidney, Craig and Berry, Dominic W. and Wiebe, Nathan and McClean, Jarrod and Paler, Alexandru and Fowler, Austin and Neven, Hartmut},
   year={2018},
   month=10 }

@article{Babbush2023,
   title={Quantum simulation of exact electron dynamics can be more efficient than classical mean-field methods},
   volume={14},
   issn={2041-1723},
   doi={10.1038/s41467-023-39024-0},
   number={1},
   journal={Nature Communications},
   publisher={Springer Science and Business Media LLC},
   author={Babbush, Ryan and Huggins, William J. and Berry, Dominic W. and Ung, Shu Fay and Zhao, Andrew and Reichman, David R. and Neven, Hartmut and Baczewski, Andrew D. and Lee, Joonho},
   year={2023},
   month={7} }

@article{Su2021,
  title = {Fault-Tolerant Quantum Simulations of Chemistry in First Quantization},
  author = {Su, Yuan and Berry, Dominic W. and Wiebe, Nathan and Rubin, Nicholas and Babbush, Ryan},
  journal = {PRX Quantum},
  volume = {2},
  issue = {4},
  pages = {040332},
  numpages = {82},
  year = {2021},
  month = {Nov},
  publisher = {American Physical Society},
  doi = {10.1103/PRXQuantum.2.040332}
}

@article{Stare2003,
    author = {Stare, Jernej and Balint-Kurti, Gabriel G.},
    title = {Fourier Grid Hamiltonian Method for Solving the Vibrational Schrödinger Equation in
Internal Coordinates: Theory and Test Applications},
    journal = {The Journal of Physical Chemistry A},
    volume = {107},
    number = {37},
    pages = {7204-7214},
    year = {2003},
    month = {09},
    issn = {1089-5639},
    doi = {10.1021/jp034440z}
}

@article{Jones2016,
   title={An efficient basis set representation for calculating electrons in molecules},
   volume={114},
   ISSN={1362-3028},
   DOI={10.1080/00268976.2016.1176262},
   number={13},
   journal={Molecular Physics},
   publisher={Informa UK Limited},
   author={Jones, Jeremiah R. and Rouet, François-Henry and Lawler, Keith V. and Vecharynski, Eugene and Ibrahim, Khaled Z. and Williams, Samuel and Abeln, Brant and Yang, Chao and McCurdy, William and Haxton, Daniel J. and Li, Xiaoye S. and Rescigno, Thomas N.},
   year={2016},
   month={4}, pages={2014-2028} }

@article{Lehtola2023,
    author = {Lehtola, Susi},
    title = {Accuracy of a Recent
Regularized Nuclear Potential},
    journal = {Journal of Chemical Theory and Computation},
    volume = {19},
    number = {13},
    pages = {4033-4039},
    year = {2023},
    month = {07},
    issn = {1549-9618},
    doi = {10.1021/acs.jctc.3c00530},
}

@article{Colbert1992,
    author = {Colbert, Daniel T. and Miller, William H.},
    title = {A novel discrete variable representation for quantum mechanical reactive scattering via the {{S}}-matrix {{Kohn}} method},
    journal = {The Journal of Chemical Physics},
    volume = {96},
    number = {3},
    pages = {1982-1991},
    year = {1992},
    month = {02},
    issn = {0021-9606},
    doi = {10.1063/1.462100},
}

@Article{Poirier2022,
   title={Full-dimensional Schrödinger wavefunction calculations using tensors and quantum computers: the Cartesian component-separated approach},
   volume={24},
   ISSN={1463-9084},
   DOI={10.1039/d1cp02036f},
   number={7},
   journal={Physical Chemistry Chemical Physics},
   publisher={Royal Society of Chemistry (RSC)},
   author={Poirier, Bill and Jerke, Jonathan},
   year={2022},
   pages={4437-4454} }

@article{Brown2019,
    author = {Brown, James and Whitfield, James D.},
    title = "{Basis set convergence of Wilson basis functions for electronic structure}",
    journal = {The Journal of Chemical Physics},
    volume = {151},
    number = {6},
    pages = {064118},
    year = {2019},
    month = {08},
    issn = {0021-9606},
    doi = {10.1063/1.5094295},
    url = {https://doi.org/10.1063/1.5094295},
}

@article{Wan2021,
  title = {Block-encoding-based quantum algorithm for linear systems with displacement structures},
  author = {Wan, Lin-Chun and Yu, Chao-Hua and Pan, Shi-Jie and Qin, Su-Juan and Gao, Fei and Wen, Qiao-Yan},
  journal = {Phys. Rev. A},
  volume = {104},
  issue = {6},
  pages = {062414},
  numpages = {17},
  year = {2021},
  month = {Dec},
  publisher = {American Physical Society},
  doi = {10.1103/PhysRevA.104.062414},
}

@article{Shang2023,
doi = {10.1088/1674-1056/acb914},
url = {https://doi.org/10.1088/1674-1056/acb914},
year = {2023},
month = {10},
publisher = {Chinese Physical Society and IOP Publishing Ltd},
volume = {32},
number = {10},
pages = {100309},
author = {Gao, Shang and Yang, Yu-Guang},
title = {A quantum algorithm for Toeplitz matrix-vector multiplication},
journal = {Chinese Physics B}
}

@article{LinTong2020,
  doi = {10.22331/q-2020-12-14-372},
  title = {Near-optimal ground state preparation},
  author = {Lin, Lin and Tong, Yu},
  journal = {{Quantum}},
  issn = {2521-327X},
  publisher = {{Verein zur F{\"{o}}rderung des Open Access Publizierens in den Quantenwissenschaften}},
  volume = {4},
  pages = {372},
  month = dec,
  year = {2020}
}

@article{White1992,
  title = {Density matrix formulation for quantum renormalization groups},
  author = {White, Steven R.},
  journal = {Phys. Rev. Lett.},
  volume = {69},
  issue = {19},
  pages = {2863--2866},
  numpages = {0},
  year = {1992},
  month = {Nov},
  publisher = {American Physical Society},
  doi = {10.1103/PhysRevLett.69.2863},
}

@article{Hohenstein2012,
    author = {Hohenstein, Edward G. and Parrish, Robert M. and Martínez, Todd J.},
    title = {Tensor hypercontraction density fitting. I. Quartic scaling second- and third-order Møller-Plesset perturbation theory},
    journal = {The Journal of Chemical Physics},
    volume = {137},
    number = {4},
    pages = {044103},
    year = {2012},
    month = {07},
    issn = {0021-9606},
    doi = {10.1063/1.4732310},
}

@article{Parrish2012,
    author = {Parrish, Robert M. and Hohenstein, Edward G. and Martínez, Todd J. and Sherrill, C. David},
    title = {Tensor hypercontraction. II. Least-squares renormalization},
    journal = {The Journal of Chemical Physics},
    volume = {137},
    number = {22},
    pages = {224106},
    year = {2012},
    month = {12},
    issn = {0021-9606},
    doi = {10.1063/1.4768233},
    url = {https://doi.org/10.1063/1.4768233},
}

@article{Lee2021,
  title = {Even More Efficient Quantum Computations of Chemistry Through Tensor Hypercontraction},
  author = {Lee, Joonho and Berry, Dominic W. and Gidney, Craig and Huggins, William J. and McClean, Jarrod R. and Wiebe, Nathan and Babbush, Ryan},
  journal = {PRX Quantum},
  volume = {2},
  issue = {3},
  pages = {030305},
  numpages = {62},
  year = {2021},
  month = {Jul},
  publisher = {American Physical Society},
  doi = {10.1103/PRXQuantum.2.030305},
}

@article{Hummel2017,
    author = {Hummel, Felix and Tsatsoulis, Theodoros and Grüneis, Andreas},
    title = {Low rank factorization of the Coulomb integrals for periodic coupled cluster theory},
    journal = {The Journal of Chemical Physics},
    volume = {146},
    number = {12},
    pages = {124105},
    year = {2017},
    month = {03},
    issn = {0021-9606},
    doi = {10.1063/1.4977994},
}

@article{Kolda2009,
author = {Kolda, Tamara G. and Bader, Brett W.},
title = {Tensor Decompositions and Applications},
journal = {SIAM Review},
volume = {51},
number = {3},
pages = {455-500},
year = {2009},
doi = {10.1137/07070111X}
}

@article{Kassal2008,
author = {Ivan Kassal  and Stephen P. Jordan  and Peter J. Love  and Masoud Mohseni  and Alán Aspuru-Guzik },
title = {Polynomial-time quantum algorithm for the simulation of chemical dynamics},
journal = {Proceedings of the National Academy of Sciences},
volume = {105},
number = {48},
pages = {18681-18686},
year = {2008},
doi = {10.1073/pnas.0808245105}}

@article{LowChuang2019,
  doi = {10.22331/q-2019-07-12-163},
  url = {https://doi.org/10.22331/q-2019-07-12-163},
  title = {Hamiltonian {S}imulation by {Q}ubitization},
  author = {Low, Guang Hao and Chuang, Isaac L.},
  journal = {{Quantum}},
  issn = {2521-327X},
  publisher = {{Verein zur F{\"{o}}rderung des Open Access Publizierens in den Quantenwissenschaften}},
  volume = {3},
  pages = {163},
  month = jul,
  year = {2019}
}

@article{Oszmian_2026,
    author = {Oszmian, Aleksandra and Lesiuk, Michał},
    title = {On the effective rank of canonical polyadic decomposition of electron repulsion integrals},
    journal = {The Journal of Chemical Physics},
    volume = {165},
    number = {6},
    pages = {064115},
    year = {2026},
    month = {08},
    issn = {0021-9606},
    doi = {10.1063/5.0343714},
}

@misc{eklund2026endtoendsimulationchemicaldynamics,
      title={End-to-End Simulation of Chemical Dynamics on a Quantum Computer}, 
      author={Elliot C. Eklund and Arkin Tikku and Patrick Sinnott and William J. Huggins and Guang Hao Low and Dominic W. Berry and Ivan Kassal},
      year={2026},
      eprint={2603.19007},
      archivePrefix={arXiv},
      primaryClass={quant-ph},
      url={https://arxiv.org/abs/2603.19007}, 
}

@article{Braess2005,
    author = {Braess, Dietrich and Hackbusch, Wolfgang},
    title = {Approximation of 1/x by exponential sums in $[1, \infty)$},
    journal = {IMA Journal of Numerical Analysis},
    volume = {25},
    number = {4},
    pages = {685-697},
    year = {2005},
    month = {10},
    issn = {0272-4979},
    doi = {10.1093/imanum/dri015},
}

@article{Beylkin2010,
title = {Approximation by exponential sums revisited},
journal = {Applied and Computational Harmonic Analysis},
volume = {28},
number = {2},
pages = {131-149},
year = {2010},
note = {Special Issue on Continuous Wavelet Transform in Memory of Jean Morlet, Part I},
issn = {1063-5203},
doi = {10.1016/j.acha.2009.08.011},
author = {Gregory Beylkin and Lucas Monzón},
}

@article{Nishiya_2024,
  title = {First-quantized adiabatic time evolution for the ground state of a many-electron system and the optimal nuclear configuration},
  author = {Nishiya, Yusuke and Nishi, Hirofumi and Couzini{\'e}, Yannick and Kosugi, Taichi and Matsushita, Yu-ichiro},
  journal = {Phys. Rev. A},
  volume = {109},
  issue = {2},
  pages = {022423},
  numpages = {10},
  year = {2024},
  month = {Feb},
  publisher = {American Physical Society},
  doi = {10.1103/PhysRevA.109.022423},
}

@misc{poirier2021efficientevaluationexponentialgaussian,
      title={Efficient Evaluation of Exponential and Gaussian Functions on a Quantum Computer}, 
      author={Bill Poirier},
      year={2021},
      eprint={2110.05653},
      archivePrefix={arXiv},
      primaryClass={quant-ph},
      url={https://arxiv.org/abs/2110.05653}, 
}

@article{schopf2014burmann,
  title={On B{\"u}rmann's theorem and its application to problems of linear and nonlinear heat transfer and diffusion},
  author={Sch{\"o}pf, Harald},
  doi={10.3888/tmj.16-11},
  journal={The Mathematica Journal},
  year={2014}
}
\bibliographystyle{apsrev4-2}

\end{document}